\documentclass{article}

\usepackage[utf8]{inputenc} 
\usepackage{amsmath} 
\usepackage{graphicx} 
\usepackage{enumitem} 
\usepackage{fancyhdr} 
\usepackage{geometry} 
\usepackage{csquotes} 
\usepackage{hyperref} 
\usepackage{apacite} 
\usepackage{booktabs} 
\usepackage{array}    

\title{Data Science Students Perspectives on Learning Analytics: An Application of Human-Led and LLM Content Analysis}
\author{Raghda Zahran, Jianfei Xu, Huizhi Liang, Matthew Forshaw \\ \textit{Newcastle University}}
\date{January 2025}

\begin{document}

\maketitle
\begin{abstract}

    \textbf{Domain:} This article contributes to the field of learning analytics within higher education, focusing specifically on insights generated by postgraduate students specialising in data science. It aims to enrich the broader discourse on relevant learning analytics in education. \\
    
    \textbf{Objective:} This study is part of a series of initiatives at a UK university designed to cultivate a deep understanding of students' perspectives on analytics resonating with their unique learning needs. It explores the results of collaborative data processing undertaken by postgraduate students who examined an Open University Learning Analytics Dataset (OULAD), thus illuminating the connections between their experiences in data science and their comprehension of learning analytics. \\

    \textbf{Methods:} A qualitative approach was adopted, integrating a Retrieval-Augmented Generation (RAG) and a Large Language Model (LLM) technique with human-led content analysis to gather information about students' perspectives based on their submitted work (conceptual understanding, formulated questions, and interpretations) of learning analytics. The study involved 72 postgraduate students in 12 groups, all enrolled in a four-week group Innovation Project as part of their module in data science. \\

    \textbf{Findings:} The analysis of group work revealed diverse insights into essential learning analytics from the students' perspectives. All participating groups adopted a structured data science methodology, employing various techniques, including linear regression analysis, influence assessments, and predictive analytics. The questions formulated by the student groups were categorised into seven themes, reflecting their specific areas of interest. Although there is variation in the selected variables to examine and interpret the correlations, a consensus was found regarding the general results. Only a limited number of groups engaged in predictive analytics, including those that used multiple models and critically assessed their levels of accuracy. \\

    \textbf{Conclusion:} A significant outcome of this study is that students specialising in data science exhibited a deeper understanding of learning analytics, effectively articulating their interests through inferences drawn from their data analyses. While a human-led content analysis enabled a general understanding of the students' perspectives, the use of LLM provided nuanced insights. Continued discussions and research focused on the scoping process and designing learning analytics and its influence on student experiences are crucial for developing a comprehensive understanding within educational contexts. \\

\end{abstract}

\section{Introduction}

Learning analytics in higher education is an evolving field focused on the measurement, collection, analysis, and reporting of data concerning university students and their contexts. A significant emphasis is placed on providing meaningful information for both students and staff, which addresses one of the primary challenges faced in higher education \cite{Educause2024}. Furthermore, fostering a culture of data-driven decision-making is essential \cite{Jisc2024}. This underscores that deriving actionable insights from this data, along with establishing robust governance structures, is a key priority for institutions seeking to enhance their educational practices and outcomes. However, such initiatives primarily depend on the careful selection of appropriate data variables and the meaningful processing of integrated data.

Despite the increasing emphasis on human- and user-centred approaches that incorporate students' perspectives \cite{Tsai2022, Martinez2023, Buckingham2024, Topali2024, Alfredo2024a}, effectively scoping and designing meaningful learning analytics remains a significant challenge for higher education institutions. Determining the scope of learning analytics necessitates a clear definition of objectives, boundaries, and analytical methodologies. The design phase involves selecting relevant data sources, preparing the data, determining appropriate variables, and effectively visualising the analytics to enhance student learning experiences and outcomes\cite{Mougiakou2023}.

Since its emergence in the early 2000s, numerous studies have explored students' perceptions of learning analytics \cite{Mcpherson2016, West2020, Divjak2023, Jovanovic2021}. These studies aim to understand the significance, limitations, challenges, opportunities, and benefits of learning analytics for students and staff, the primary users and key stakeholders. However, most studies have identified a challenge in obtaining students' perspectives, primarily due to their limited conceptual understanding of learning analytics. This highlights the importance of incorporating students' attitudes and feedback on design, underscoring the need for an analytics-aware, student-centred approach to developing effective learning analytics systems.

While some research has successfully engaged students, few studies have established a clear connection between their understanding and the potential applications that may emerge from this synthesis. Investigations into students' perceptions reveal a significant gap in mechanisms that encourage a shift in student thinking—from merely learning processes to considering the epistemological implications of data related to learning.

Studies examining student engagement highlight the importance of experiences in shaping perspectives on effective teaching and learning. Data- and analytics-aware student insights are particularly valuable as they evolve throughout their academic journeys. This underscores the need for a collaborative and inclusive design in learning analytics that accommodates the diverse needs of the student body. Such an approach not only empowers student voices but also fosters the development of relevant and impactful learning strategies that reflect their unique contexts.

Emerging collaborative design (Co-design) studies promote collective views among participating students. Co-design is described as a way of working that brings together different stakeholders—including end-users, designers from an 'expert perspective', and other relevant parties—to jointly create solutions that are more relevant and effective \cite{Sanders2008}. Some studies have concluded that adopting a collaborative approach to engaging students in learning design offers personal benefits, such as enhancing their confidence, engagement, deep understanding, and knowledge development. In addition, it promotes social benefits, including interaction and collaboration between students.

In line with this collaborative ethos \cite{Sanders2008}, this study outlines our method for gathering the perspectives of postgraduate students specialising in data science as part of their innovation project. We aim to gain a deeper understanding of students' viewpoints through two key strands:

\begin{itemize}
    \item \textbf{Postgraduate Students:} These experienced learners synthesise information and develop innovative designs to tackle specific challenges. They possess a comprehensive understanding of the learning process and can pinpoint the data and analytics that would enhance their educational experience.
    
    \item \textbf{Data Science Specialists:} This strand explores the role of students as data science experts, where they recognise the importance of analytics and employ various techniques to extract insights from data, transforming these into meaningful visual representations.
\end{itemize}

By integrating these perspectives, we seek to enrich the understanding of learning analytics for both students and educational institutions. This study aims to investigate how students articulate their insights through their design outputs, focusing specifically on three research questions:

\begin{itemize}
    \item How does students' work reflect their understanding and application of relevant learning analytics?
    \item What specific questions and variables do data science students articulate in their learning analytics co-design and development process?
    \item What insights and interpretations related to learning analytics do data science students derive through their co-design and development activities?
\end{itemize}

By investigating these questions, this study aims to illuminate the connections between students' practical experiences in data science and their conceptual understanding of learning analytics, thereby enriching the broader discourse on relevant learning in education.

The subsequent section reviews recent literature related to learning analytics and student-centred approaches. Following this, the methodology section details our content analysis, which integrates Retrieval-Augmented Generation (RAG) and large language models (LLM) to examine the outputs of students' innovation projects. Our analytical framework combines both human and LLM-based methods, offering an innovative approach to evaluating student perspectives, and designs and generating collective insights from their work. This strategy effectively leverages the strengths of traditional information retrieval systems, such as databases.

The findings section provides valuable information about students' perspectives, highlighting their approaches, formulated questions, and interpretations. Finally, the section on limitations and recommendations delineates the boundaries of this study and offers our perspectives on future research directions.

\section{Previous Evaluation of Students' Perspectives}

The literature on learning analytics has undergone a significant transformation, increasingly prioritising student-centred approaches. This shift underscores the importance of designing software tools and processes that focus on the needs and experiences of users \cite{Alfredo2024b, Topali2024, Buckingham2024}. Learning analytics has emerged as a promising framework to better support and understand students' learning processes \cite{Schumacher2018}.

\subsection{Human-Centred Design}
A key aspect of this evolution is the recognition that effective learning analytics must employ interdisciplinary methodologies, incorporating insights from human-computer interaction (HCI), educational design, and software engineering \cite{Martinez2016, Dollinger2018, Revano2021, Tsai2022}. Designing analytics with users, rather than for them, presents challenges related to the specialised skills that users may lack. Therefore, a collaborative approach is essential, ensuring that designers work closely with students and educators to develop tools that genuinely enhance learning experiences.

\subsection{Frameworks and Models}
Recent studies examining student perspectives have highlighted the significance of frameworks such as Learning Design Analytics Layers (AL4LD) \cite{Hernandez2019} and the Learning Analytics Model (LAM) \cite{Defreitas2015}. These frameworks exemplify innovative approaches that inform learning design decisions, contributing to the emerging field of human-centred learning analytics (HCLA) that prioritises student needs.

The adoption of human-centred design principles (HCD) is crucial for engaging students effectively. However, implementing these principles presents challenges in creating actionable learning analytics solutions that balance stakeholder agency with the often technocratic tendencies of data science \cite{Martinez2023}. By prioritising student engagement, educators can develop more responsive analytics that enhance the learning experience.

\subsection{Stakeholder Engagement}
Engaging stakeholders, including students who may be generalists but adept in analytics, is vital \cite{Martinez2023}. Although their insights into learning processes are valuable, many studies indicate a lack of conceptual understanding of analytics among these individuals \cite{Mcpherson2016, Gasevic2019, Divjak2023}. This gap highlights the need for deeper engagement with learning analytics principles to maximise effectiveness in educational contexts.

Integrating user-informed learning analytics into educational environments is expected to facilitate timely feedback and improve the overall learning experience. This capability is particularly crucial in technology-enhanced settings, where data access can guide instructors and students in making informed decisions about learning strategies \cite{Schmitz2018, Saint2018, Dimitriadis2021}. Evidence suggests that effective feedback mechanisms can significantly boost student engagement and retention, empowering learners to take control of their educational journeys \cite{Jovanovic2021, Bodily2018, Wang2021}.

\subsection{Evidence-Based Practices}
The application of evidence-based practices is increasingly relevant in educational contexts. Learning analytics holds the potential to generate actionable insights that inform teaching practices and enhance student success \cite{Kuromiya2020}. Recent systematic reviews highlight learning analytics' capacity to improve feedback mechanisms in higher education, bridging the gap between data collection and pedagogical application \cite{Ifenthaler2020, Sahin2021, Banihashem2022}.

\subsection{Challenges and Ethical Considerations}
Despite the acknowledged potential of learning analytics to transform higher education, several challenges remain, including resource allocation, stakeholder participation, and ethical considerations \cite{Tsai2019, West2020, Joseph2021}. Addressing these issues is critical for institutions seeking to fully realise the benefits of learning analytics, such as enhanced teaching and improved administrative efficiencies.

As the field evolves, the successful implementation of learning analytics increasingly relies on agile leadership capable of navigating environmental pressures and managing conflicts \cite{Tsai2019}. Furthermore, understanding students' perceptions of data privacy and the ethical implications of analytics practices is essential for fostering trust and ensuring respect for personal and professional boundaries \cite{West2020}.

However, an overemphasis on data-driven processes—often dominated by data scientists and technologists—can obscure the nuanced social dimensions of educational contexts. A socio-technical approach is essential for developing and implementing effective learning analytics solutions, considering the interplay of human factors to achieve holistic educational outcomes.

\subsection{The Role of Artificial Intelligence}
Recent studies have begun to explore the integration of artificial intelligence (AI) within student-centred design frameworks, highlighting its potential to transform educational settings \cite{Alfredo2024a, Topali2024}. AI technologies, including machine learning and natural language processing, are increasingly recognised as tools for analysing student-generated content and enhancing learning experiences \cite{Samadi2024}.

The literature underscores the growing significance of AI in education, particularly in learning analytics, feedback systems, and engagement analysis. AI-driven tools have demonstrated the ability to improve collaborative learning assessments by providing real-time insights into group dynamics and individual contributions \cite{Pradhan2024}. Additionally, automating feedback processes through AI can yield timely and personalised responses, which are crucial for student engagement and motivation. Addressing algorithmic biases in AI systems is vital for promoting educational equity, ensuring that all students benefit from these technologies \cite{Asatryan2024}.

The integration of Large Language Models (LLMs) is also being explored for their capacity to personalise learning experiences and refine assessment methodologies. These models can adapt to individual learning styles and preferences, thus enhancing the overall educational experience \cite{Rashid2024}. This potential for personalisation is particularly important as educational environments become increasingly diverse.

As student demographics evolve and enrolment numbers increase, there is a pressing need for dynamic analytics approaches that can accommodate a larger and more varied student population. Strategies aligned with universal design principles \cite{Cast2024} can ensure that analytics tools are accessible and effective for all learners, thereby enriching the educational experience.

Research indicates that AI can assist educators in understanding the needs of individual students, fostering a more responsive learning environment. As the body of literature on this topic expands, it becomes evident that AI has the potential to significantly enhance the analysis of student interactions and outcomes.

\subsection*{Limitations and Gaps in the Literature}

Despite the advancements in learning analytics literature, significant limitations and gaps warrant further exploration. One major limitation is the technocentric design of many learning analytics tools, which often neglects the diverse needs and perspectives of students and educators. This oversight can hinder their effectiveness in real-world educational contexts. Furthermore, while many studies emphasise personalisation and real-time feedback, there is a lack of inclusive design that integrates these elements into a cohesive approach. Ethical concerns regarding data privacy, informed consent, and transparency of analytics practices are also inadequately addressed, raising questions about the trustworthiness of these systems.

Moreover, the literature often focuses on descriptive studies rather than experimental or longitudinal research, limiting the understanding of the long-term impact of learning analytics on student outcomes. There is a pressing need for empirical investigations into the effectiveness of specific learning analytics interventions, particularly in diverse disciplinary contexts, to better inform the design and implementation of these systems. The gap in participation among stakeholders who are generalists or new to learning analytics highlights the importance of developing data and analytics-aware skills among students and staff, as their contributions are crucial for the successful implementation of learning analytics practices.

To address these gaps in engagement, this study focuses on integrating data science students through a participatory co-design project-based approach. By involving students with data science expertise and transferable skills, the study aims to foster deeper collaboration between students and educators. This approach enhances the educational experience and empowers students to actively contribute to the analytics process, ensuring that the developed tools are user-friendly and grounded in a solid understanding of both data and the learning environment. As institutions prepare to embrace this new paradigm, the involvement of students with dual competencies will be vital in designing effective, human-centred learning analytics solutions.

\section{Method}
This study employs a qualitative approach that integrates human-led content analysis with retrieval-augmented generation (RAG) and large language models (LLMs). Qualitative Content Analysis (QCA) is a method for drawing inferences from texts and other meaningful materials, allowing researchers to interpret their findings through a structured process \cite{Bengtsson2016, Krippendorff2018, Marvasti2019, Silverman2024}. To enrich our analysis further, we incorporate a quantitative perspective, enabling the systematic identification of emerging themes.

\subsection*{Study Design}
This study utilises content analysis to examine the themes present in the objectives, interpretations, tools, and techniques found in student reports. We validate our findings through replicated and diverse analyses. The combination of retrieval-augmented generation (RAG) and large language models (LLMs) aims to enhance the precision and relevance of the generated content \cite{Dai2023, Arslan2024}, facilitating a deeper understanding of the data itself. Ultimately, our goal is to provide nuanced insights into the students' learning experiences, illuminating the complexities of their educational journeys.

The overarching qualitative approach is enriched by examining trends within the categories and themes that resonate with students' interests. This is complemented by a quantitative analysis of these themes to assess the degree of student engagement. Ultimately, our objective is to synthesise the diverse methodologies employed by the students, thereby capturing their perspectives within the context of their university experiences.

\subsection*{Participants}
The study involved 72 postgraduate students enrolled in a four-week group project as part of a module offered to different cohorts throughout the academic year. This module aimed to deepen the students' understanding of the processes and skills essential for developing a data science system through hands-on application in collaborative innovation. Recognising these students as experienced in data science aligns with findings indicating that prior knowledge enhances their contributions to discussions.

Viewing students as experienced informants enriches this educational research \cite{Cilesiz2011}. In Participatory Action Research (PAR), students draw on their knowledge, ensuring that findings reflect real-world contexts \cite{Reason2008}. Their participation in case studies also adds vital context, improving the analysis \cite{Yin2018}. Therefore, acknowledging postgraduate students as experienced in learning processes and educational technology \cite{Cilesiz2011} fosters a more inclusive educational systems design.

This cohort was organised into 12 small groups, each supported by a module demonstrator who provided guidance on technical aspects and addressed administrative requirements. This approach highlights the importance of facilitation and support in group work.

\subsection*{Ethical Considerations}
The participating students were informed that their submissions would be used to investigate their views on effective learning analytics. The collected data from students' reports were anonymised, ensuring no possibility of linking students' identities to their submissions. Participants were made aware that the insights gained would be utilised for scoping and subsequent reporting, thereby fostering an environment of trust and accountability throughout the exploration process.

\subsection*{Content Analysis}
To gain a deeper understanding of the participating students' perspectives, we employed a Qualitative Content Analysis of Learning Analytics (QCALA). This approach emphasises qualitative methods within the realm of learning analytics and is adapted from the content analysis process proposed by Marvasti \cite{Marvasti2019}. In addition to this foundation, our methodology integrates both human analysis and Large Language Models (LLMs) analysis \cite{Zhang2023}. By adopting this dual approach, we aim to capture richer insights into the nuances of student engagement and analytical thinking.

To operationalise QCALA, we followed a systematic process consisting of several key steps:

\begin{enumerate}[label=\textbf{\arabic*.}]
    \item \textbf{Defining the Research Aim:} 
    The authors concurred that the primary objective was to identify analytics that resonate with university students and effectively address their learning needs, particularly from the perspective of those who have developed a conceptual understanding of data science and analytical methods. This aim was clearly articulated to the participants through a well-defined problem statement that outlined the research objectives and initial requirements at the outset of the module.

    \item \textbf{Source Material:} 
    The reports produced by the participating students presented their work in response to a specified dataset within a module design context. The students were tasked with utilising the anonymised Open University Learning Analytics Dataset (OULAD) \cite{Kuzilek2017} to explore and design learning analytics from their perspectives, aligning well with the data scope of their home institution. The open-source nature of the dataset facilitated student access via an online link, while the anonymisation process was implemented to mitigate privacy risks and ensure compliance with data protection regulations. Although the dataset originated from another higher education institution, it was rigorously examined by the authors to ensure it met the requirements of the students' home university. This methodology aligns with Marvasti's \cite{Marvasti2019} principles of employing relevant documents as data sources and Bengtsson's \cite{Bengtsson2016} idea of selecting units of analysis that inspire sufficient confidence.
     
    \item \textbf{Sample Documents:} 
    The authors opted to analyse the entire set of reports rather than a sample. This decision allows for a thorough and nuanced understanding of the entire cohort's perspectives, aligning with Krippendorff's \cite{Krippendorff2018} assertion that comprehensive analysis enhances the interpretive context. It not only reduces the risk of sampling bias but also reflects a commitment to rigour and meticulousness in line with Marvasti's principles \cite{Marvasti2019}. Furthermore, our approach can be likened to cluster sampling, wherein the entire group serves as the unit of analysis. This is appropriate given the finite and manageable number of available reports, allowing for a more robust exploration of the data.

    \item \textbf{Identifying Categories or Features:} 
    To construct a collective understanding of student perspectives, the focus was placed on both textual and visual outcomes \cite{Marvasti2019}. The first and second authors examined theoretical underpinnings and frameworks, formulated questions, and interpretations by students to explore the given problem statement and the objectives of their work, mirroring Marvasti's \cite{Marvasti2019} recommendation to pinpoint key themes and categories relevant to the research problem. This analysis revealed the students' interests and focal points, as well as the methods, tools, and techniques they selected, reflecting their diverse approaches to data science. Furthermore, their overall reflections and written interpretations \cite{Silverman2024} highlighted perceived issues and challenges encountered during the process, representing their lived experiences and profound perspectives \cite{Cilesiz2011}. The visual analysis scrutinised technical artefacts, such as dashboards and diagrams, employing semiotic analysis to develop nuanced perspectives among students, which is essential for a comprehensive understanding.     

    \item \textbf{Measure Count:} 
    Our qualitative analysis was followed by a quantitative assessment of the frequency of these themes, which is recommended in content analysis \cite{Bengtsson2016, Silverman2024}. This dual approach permits a nuanced understanding of the data, aligning with Marvasti's method of combining qualitative insights with quantitative measures. The analysis focused on the outcomes, contextualised within the course material, since student work was undertaken within a semi-controlled environment featuring a predetermined dataset and overarching objectives. Additionally, we intentionally excluded demographic information, social aspects of the student experiences, and their verbal reflections, although these elements are essential for interpretation \cite{Silverman2024}, as they have been explored in other university initiatives concerning the development of learning analytics.
    
    \item \textbf{Employing LLM-RAG:} 
    In addition to the human analysis, the first and third authors manually examined the content of student reports through thematic analysis of key elements, formulating questions, interpretations, and the tools and techniques employed. The second author utilised a Large Language Model (LLM) with Retrieval-Augmented Generation (RAG) \cite{Lewis2020} to conduct the analysis. This study employs GPT-4 as the generative model, integrated within an RAG system framework \cite{Gao2023}. The system extracts content from PDF files using PyMuPDFReader \cite{Pymupdf2025} and converts report content into fixed-length sentence embeddings using a pre-trained model (sentence-transformers/all-distilroberta-v1) \cite{Huggingface2025}. These embeddings are stored in the open-source Weaviate vector database as an external knowledge repository for the LLM. When users interact with the model by posing questions, the model utilises the HNSW (Hierarchical Navigable Small World) \cite{Malkov2020} retriever to fetch relevant information from the external knowledge repository. It selects the top 50 most relevant chunks of contextual information and leverages GPT-4's capabilities to generate answers for question-answering tasks in learning analytics.
\end{enumerate}

\begin{figure}
    \centering
    \includegraphics[width=1\linewidth]{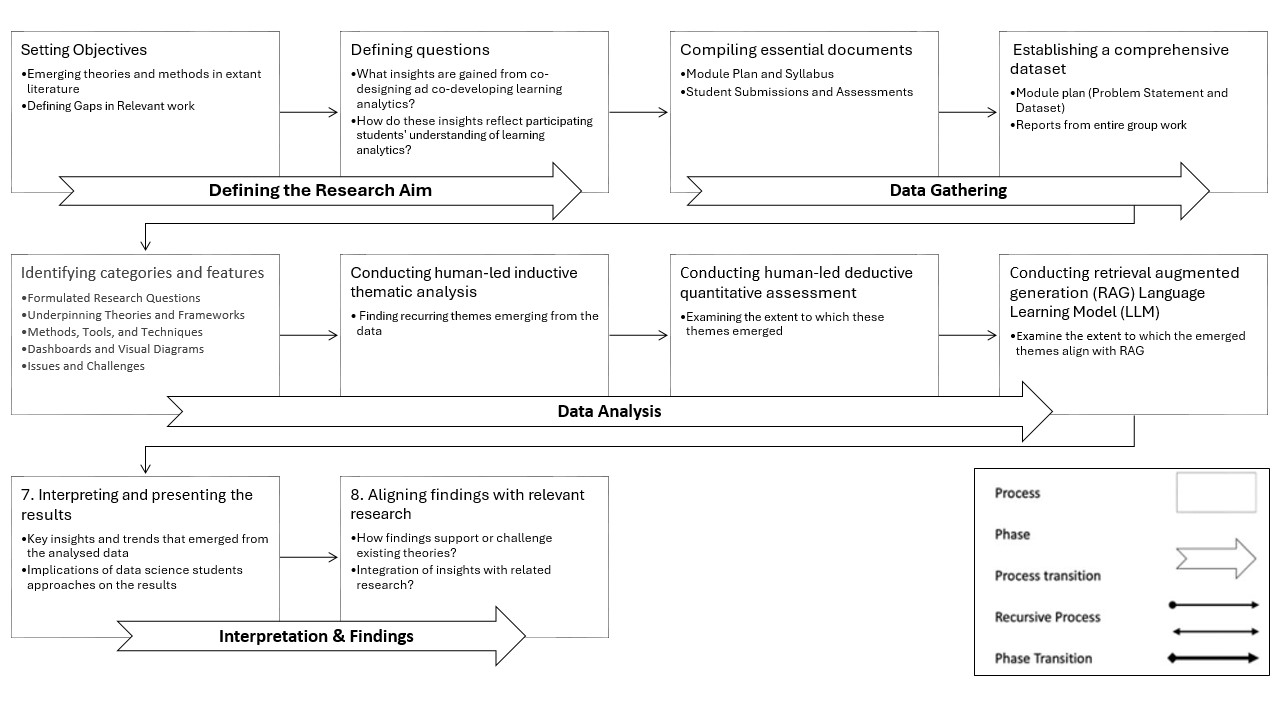}
    \caption{Student Perspectives: Qualitative Content Analysis of Learning Analytics (QCALA) Process}
    \label{fig:figure_1_student_perspectives_qualitative_learning_analytics_contentanalysis_process}
\end{figure}

\section{Findings Based on Human Analysis}

In this study, we aimed to address gaps concerning student engagement in the design of learning analytics services. Recent studies characterised by human-centred learning analytics design have predominantly involved generalist students from various disciplines, utilising structured approaches to examine their perceptions of learning analytics. In contrast, this study delves deeper into the perspectives of students with diverse experiences, shaped by their backgrounds as postgraduate students and specialists in data science.

\subsection{Student Conceptual Understanding}

Our first research question was: \textbf{\textit{How does students' work reflect their understanding and application of relevant learning analytics?}} By adopting a structured approach to data mining, the questions and conclusions formulated by the groups indicated a profound understanding of the aims of learning analytics and the data derivable from available datasets concerning student engagement and achievement. However, our analysis revealed that while most groups adhered to the Cross-Industry Standard Process for Data Mining (CRISP-DM) framework—which consists of the phases: business understanding, data understanding, data preparation, modelling, evaluation, and deployment—their work lacked grounding in the existing learning analytics literature. This absence could have provided them with crucial direction to formulate their questions and interpretations, thereby enriching their understanding.

Implementing the CRISP-DM framework, the groups' data preparation involved the utilisation of the dataset by selecting relevant data for analysis and formatting it appropriately. Key steps included:

\begin{itemize}
    \item \textbf{Outlier Detection:} Identifying and addressing outliers to ensure data integrity, such as excluding outliers like the course or module home page from calculations of engagement or extreme grades.
    \item \textbf{Filtering Confounding Variables:} Removing variables that could distort the analysis.
    \item \textbf{Reducing Duplicates:} Ensuring data uniqueness to enhance accuracy.
    \item \textbf{Addressing Missing Values:} Employing methods such as imputation to fill gaps in the dataset.
\end{itemize}

Additionally, categorical variables were encoded using label encoding, and perfect collinearity was eliminated to refine the dataset further. Statistical calculations were performed on assessment scores and virtual learning environment (VLE) interactions, including:

\begin{itemize}
    \item \textbf{Engagement Metrics:} The sum was used to calculate the overall level of engagement.
    \item \textbf{Assessment Metrics:} Sums, averages, means, and modes were utilised to assess performance at both the individual student level and across multiple cohorts.
\end{itemize}

These steps not only showcased the groups' data science skills but also provided a robust foundation for further analysis and visualisation. The application of data science methodologies afforded insights that the groups deemed valuable, guiding decision-making processes and highlighting areas for further investigation. This ensured that their analytical efforts had a meaningful impact on business objectives.

Most groups conducted exploratory data analysis (EDA) by establishing objectives through formulated questions or hypotheses, subsequently developing models using technological tools such as R Studio, Python libraries, and Power BI. Each group employed specific analytical techniques and utilised various visualisations. Collectively, they applied a diverse range of methods, including two-dimensional (2D) and three-dimensional (3D) models, regression analysis to examine correlations between variables, influence assessments to investigate causality, predictive analytics based on historical data, and distribution models to understand the variances within the selected dataset.

It is suggested that the pragmatic nature of the course and project contributed to this limitation, resulting in a cautious understanding where students recognised that the data could be indicative rather than definitive due to a lack of contextualisation. They acknowledged that their findings could serve as a foundation for further exploration through direct engagement with peers and faculty, thereby supporting student learning design and other services.

\subsection{Student Collaborative Design and Development}

This section explores the collaborative design and development processes of data science students in the context of learning analytics, guided by the research question:\textbf{\textit{What specific questions and variables do data science students articulate in their learning analytics co-design and development process?}} This inquiry consists of two primary strands: the students' co-design approach and the specific questions and variables they formulate, reflecting their interests and perspectives on useful analytics.

The innovative course design, alongside a group innovation project, encourages students to engage in co-designing and developing their analytical perspectives. While this structure is intended to promote a collective viewpoint, the reality of students working independently outside the classroom raises questions about the depth of their collaboration and individual contributions. Measures have been implemented to mitigate collaboration bias, including direct supervision by administrators and the requirement for individual reflective reports. However, examining the effectiveness of these measures fell outside the scope of this study.

To investigate the second strand, a thematic analysis was conducted on the questions formulated by the students. This analysis revealed that the insights generated by data science students effectively addressed pressing questions pertinent to their areas of interest. However, the questions tended to be limited in scope, lacking overarching targets and depth that could yield more insightful findings. Even though the groups examined the same dataset, their conclusions varied, exposing both similarities and contradictions in their analyses.

The inquiries posed by the students illustrate their interest in exploring multiple dimensions of performance, including assessments, engagement, and personal factors such as disabilities. This holistic approach allows for a comprehensive understanding of how various elements interact in the educational environment. The language used in formulating these questions reflects the students' data science training, with a strong emphasis on measurable outcomes that can be investigated through statistical methods. Their inquiries incorporate specific terminology and concepts, including relationships, impact, effect, and prediction, thereby establishing a more technical framework for analysing educational data, distinguishing this approach from more traditional studies.

To guide their research objectives, some groups formulated overarching questions, such as: “Do students have a higher chance of success the more they interact with Newcastle University’s educational technology?” While most groups generated fragmented questions, the primary questions often led to several sub-questions aimed at achieving their research objectives. Some analyses prompted the formulation of additional questions, indicating that the process was exploratory and illuminated areas of uncertainty and complexity that merit ongoing investigation.

While the selected variables and questions serve as indicators of students’ interests, our primary focus was on the themes that emerged and their representation across inquiries, offering insights into broader educational concerns and priorities. By comparing these themes, we gained a deeper understanding of students’ perspectives on performance and engagement.

A significant emphasis in the questions is on engagement, not merely as a metric but as a crucial factor influencing course outcomes. This suggests a need for deeper exploration into how student involvement affects learning success. Some sub-questions aim to identify predictive relationships, such as how assessment scores and engagement levels can forecast overall course performance. This forward-looking perspective is often less emphasised in conventional educational assessments.

The inclusion of questions addressing the impact of resits and disabilities introduces a vital dimension of equity and inclusivity, recognising that diverse student backgrounds and challenges significantly influence educational outcomes. Furthermore, the consideration of contextual factors—such as course type, duration, and timing of interactions—highlights the importance of situational variables in educational results, often overlooked in broader studies.

Overall, the predominant themes identified across the formulated questions include analysing the relationship between virtual learning environment (VLE) engagement and student performance, understanding the impact of assessment structures, and identifying at-risk students. The variation in scope, detail, and types of analytics employed—spanning descriptive and predictive analytics—illustrates the diversity of approaches taken. There were attempts at diagnostic analytics to examine the influence of variables on one another, although these efforts were not as pronounced. The absence of more robust diagnostic analytics indicates a potential gap in comprehending the underlying causes of student performance issues, suggesting that such analyses may require a more nuanced approach. Addressing this gap could provide valuable insights for developing targeted interventions and enhancing student support. A more comprehensive approach to question formulation could ultimately offer a richer understanding of the complexities involved in student engagement and performance.

In conclusion, the collaborative process undertaken by data science students reveals both the potential and the limitations inherent in their analyses. While the engagement with learning analytics is promising, further refinement of their approaches could yield deeper insights and more effective strategies for enhancing educational outcomes.

Table 1 summarises these themes and illustrates the varying levels of interest based on the questions formulated by the different groups.

\begin{table}[htbp]
    \centering
    \small 
    \caption{Levels of Interest Based on Questions and Groups}
    \resizebox{\textwidth}{!}{ 
    \begin{tabular}{@{}p{4cm}p{3cm}p{3cm}p{0.5\textwidth}@{}}
        \toprule
        \textbf{\raggedright Category/Theme} & \textbf{\raggedright Interest Level (Questions)} & \textbf{\raggedright Interest Level (Groups)} & \textbf{\raggedright Students' Formulated Questions} \\
        \midrule
        Assessment and Attainment & 27\% & 58\% & 
        \begin{tabular}[c]{@{}p{0.5\textwidth}@{}} 
            1. What is the relationship between assessment scores and course scores? \\ 
            2. How do assessment plans affect overall scores? \\ 
            3. What is the link between the number of assessments and scores? \\ 
            4. How does assessment type impact student performance? \\ 
            5. What do initial exam results indicate? \\ 
            6. What are the pass rates for resits? \\ 
            7. How can we predict a student's likelihood to take an exam? \\ 
            8. How does student performance progress over time? \\ 
            9. How do assessment results predict course scores?
        \end{tabular} \\ 
        \midrule
        
        Engagement and Attainment & 15\% & 67\% & 
        \begin{tabular}[c]{@{}p{0.5\textwidth}@{}} 
            1. What is the relationship between VLE engagement and assessment scores? \\ 
            2. How do specific activities affect overall course scores? \\ 
            3. What is the impact of engagement with assessments on final scores? \\ 
            4. How does the number of assessments relate to engagement? \\ 
            5. Can we predict course scores based on engagement?
        \end{tabular} \\ 
        \midrule

        Subject, Discipline, Performance, Engagement & 9\% & 17\% & 
        \begin{tabular}[c]{@{}p{0.5\textwidth}@{}} 
            1. How does module subject relate to overall scores? \\ 
            2. What is the relationship between discipline and course scores? \\ 
            3. How does discipline impact overall engagement?
        \end{tabular} \\ 
        \midrule

        Student Characteristics and Outcomes & 9\% & 33\% & 
        \begin{tabular}[c]{@{}p{0.5\textwidth}@{}} 
            1. What does the withdrawal of disabled students indicate? \\ 
            2. How does disability influence student results? \\ 
            3. Can we predict disability based on engagement and results?
        \end{tabular} \\ 
        \midrule

        Engagement Metrics & 15\% & 50\% & 
        \begin{tabular}[c]{@{}p{0.5\textwidth}@{}} 
            1. What is the level of engagement in courses? \\ 
            2. How do students engage with specific activities? \\ 
            3. What are the engagement levels across different courses? \\ 
            4. How does course type relate to engagement? \\ 
            5. Can we predict overall engagement from activity types?
        \end{tabular} \\ 
        \midrule

        Temporal Factors & 12\% & 58\% & 
        \begin{tabular}[c]{@{}p{0.5\textwidth}@{}} 
            1. How does the timing of assessments affect engagement and scores? \\ 
            2. What is the impact of module duration on engagement? \\ 
            3. How does course duration affect overall scores? \\ 
            4. What engagement trends occur around assessment due dates?
        \end{tabular} \\ 
        \midrule

        Retention and Withdrawal & 12\% & 33\% & 
        \begin{tabular}[c]{@{}p{0.5\textwidth}@{}} 
            1. When do students typically withdraw from courses? \\ 
            2. What are the withdrawal rates across different courses? \\ 
            3. How does engagement relate to withdrawal? \\ 
            4. What is the connection between engagement and course completion?
        \end{tabular} \\ 
        \bottomrule
    \end{tabular}
    } 
\end{table}

\subsection{Student Interpretations and Insights}
While similarities existed in the questions formulated by students, their interpretations diverged significantly. The reflections in their reports demonstrated a keen enthusiasm for exploring the data and articulating insights. Each group employed structured approaches to interpret the data and analytics, selecting variables and providing reasoned justifications for their choices. Overall, the analyses were specific to the dataset, resulting in varied interpretations among groups.

Depending on the selected subset of variables and courses, students' interpretations reflected a diverse range of possibilities. For instance, some groups concluded that certain variables were correlated, while others contested these findings. Furthermore, some groups conducted additional analyses, discovering correlations within specific subsets of data that did not exist in others. This indicates that a more detailed examination could yield nuanced insights.

Table 2 summarises the key variables identified by the students, along with their respective interpretations and correlations.

\begin{table}[htbp]
    \centering
    \caption{Selected Variables and Their Correlations}
    \begin{tabular}{@{}lll@{}}
        \toprule
        \textbf{Variable}                       & \textbf{Interpretation}                                   & \textbf{Correlation} \\ \midrule
        Engagement Factors                     & Influence on coursework relations and time management    & Varies by course     \\
        Final Scores                           & Impact of module length and assessment results            & Negative             \\
        Engagement and Student Success         & Active participation correlates with performance          & Positive             \\
        Temporal Engagement                    & Variability in motivation throughout the semester         & Fluctuating          \\
        Specific Engagement Elements            & Interest in activity types and assessment dates          & Context-dependent     \\
        Challenges for Disabled Students       & Higher withdrawal rates due to learning challenges        & Negative             \\
        Submission Behaviour                   & Early submissions linked to better retention              & Positive             \\
        Resits and Retention                   & Relationship between resits and engagement                & Positive             \\
        Overall Performance                    & Factors affecting overall scores and distinctions         & Positive             \\
        Regular Assessments                    & Influence on engagement and continuous learning           & Positive             \\ \bottomrule
    \end{tabular}
\end{table}

Most groups inferred that the data alone cannot provide definitive correlations or causal relationships due to the complexities of students' contexts. They highlighted the existence of external variables not included in the dataset, which could confound their interpretations and enhance understanding. Most conclusions were contextual and speculative, indicating a level of uncertainty. Some groups displayed greater confidence in the relationships and influences between variables based on statistical significance than others.

Several groups identified insights beyond the established objectives. For instance, one group discovered that longer courses were associated with higher dropout rates and that submission timing influenced final scores. This finding underscores the students' critical thinking abilities and the dynamic, contextual nature of learning analytics. Additionally, several groups suggested that analytics could be further enriched through targeted surveys aimed at exploring their findings in greater depth.

\subsection{Insights from the QCALA Analysis}

The QCALA uncovered diverse insights into students' perspectives on effective learning analytics. By employing a dual-channel analysis of human and model approaches, we identified key commonalities and unique insights, which are presented below:

\subsubsection{Common Findings}

\begin{itemize}
    \item Both human and model analysis approaches demonstrate that the groups focus on examining student behaviours and interactions within educational contexts. They leverage learning analytics to uncover patterns and trends.
    \item There is an emphasis on exploring the relationship between engagement metrics and academic outcomes, aiming to identify key performance drivers and address challenges in connecting behaviours to measurable results.
    \item Both approaches employed robust data analysis techniques such as Exploratory Data Analysis (EDA) within frameworks like CRISP-DM and utilized tools like Python to deliver actionable insights.
\end{itemize}

\subsubsection{Insights from Human Analysis}

\begin{itemize}
    \item Human analysis reveals that the groups widely utilise the CRISP-DM framework but do not delve into its detailed implementation methods.
    \item A positive correlation between engagement with Virtual Learning Environments (VLEs) and improved performance was highlighted, noting that increased clicks are linked to better academic outcomes.
    \item Patterns such as the connection between early assignment submissions and higher scores were identified, though finer distinctions—like the impact of submission timing on different types of assessments—were not explored.
\end{itemize}

\subsubsection{Insights from RAG LLM Model Analysis}

\begin{itemize}
    \item The findings from model analysis build upon the results of human analysis by offering more detailed and nuanced insights.
    \item Model analysis validates the application of the CRISP-DM framework, exploring its components in greater depth, including data understanding, preparation, modelling, evaluation, and deployment.
    \item In-depth exploration of engagement metrics revealed specific correlations, such as the relationship between the number of clicks and outcomes like pass rates, distinctions, and withdrawals across various modules.
    \item The timing of submissions was examined, showing that students who submit assessments within the first 250 days of a course generally achieve better results than those submitting later.
    \item Additionally, the analysis indicates differences in performance between CMAs (Computer Marked Assignments), TMAs (Tutor Marked Assignments), and final exams, with TMAs often proving to be the most reliable predictor of success.
\end{itemize}

\subsubsection{Conclusion of Insights}

While the findings from human analysis provide a strong foundational understanding, they sometimes lack the detail present in model analysis. Conversely, model analysis offers richer context but may lack the succinctness of human insights. Together, these findings create a comprehensive view of student engagement and learning analytics, highlighting the importance of integrating both approaches for a more holistic understanding.

\section{Limitations}

This study is subject to several limitations that should be acknowledged:
\begin{itemize}
    \item \textbf{Timescale of the Innovation Project:} The research was conducted within the constrained timeframe of Semester 2 of the Academic Year 2022, which may have impacted the depth and breadth of the findings.
    
    \item \textbf{Semi-Controlled Environment:} The study was conducted in a semi-controlled setting, characterised by specific time constraints, predetermined problem statements, and defined objectives, along with reliance on an open-source dataset and available resources. This context may limit the generalisability of the results.
    
    \item \textbf{Focus on Outcomes:} The emphasis on measurable outcomes has led to a deliberate exclusion of contextual factors and social aspects that could provide a more nuanced understanding of the learning environment and its dynamics.
\end{itemize}

\section{Conclusion}

This study aims to illuminate the connections between students' practical experiences in data science and their conceptual understanding of learning analytics, ultimately contributing to the broader discourse on meaningful learning in education. The findings indicate that students with specialised skills are capable of providing informed insights into service design. By adopting a structured approach, critically selecting variables, and employing effective analysis and visualisation tools, students demonstrate their ability to generate valuable insights throughout the service design and development process. Furthermore, the groups' methods for formulating questions and deriving meaning from the data suggest that analytics involve an exploratory and inductive approach to understanding students' experiences.

The methods employed in this study included a combination of qualitative and quantitative analyses, utilising both human analysis and RAG LLM model analysis. The human analysis provided foundational insights through group discussions and data exploration, while the model analysis offered more nuanced, data-driven perspectives. This mixed-methods approach enabled a comprehensive understanding of student engagement and performance, integrating subjective interpretations with objective metrics. By leveraging these diverse methodologies, the study enriches the overall narrative of learning analytics.

Overall, these results highlight the necessity for continuous dialogue around analytics to foster a shared understanding between students and university staff. Such conversations can explore tailored approaches that enhance the learning experience, ensuring that both students and educators benefit from a collaborative environment. By actively engaging in these discussions, institutions can better support student success and promote meaningful learning outcomes in the ever-evolving landscape of education.

\section{Acknowledgements}
The authors wish to express their sincere gratitude to the Data Science postgraduate students for their invaluable insights into learning analytics. We also extend our heartfelt thanks to the module demonstrators for their support and guidance to the students throughout the learning process. Special thanks are due to Jim Kean, Principal Technical Consultant (Data) at Jisc, and Carina Buckly, Instructional Design Manager at Southampton Solent University, for their engaging discussions and thorough reviews. Their contributions have significantly enriched the development of this work.

\newpage

\bibliographystyle{apacite}
\bibliography{references}

\end{document}